\documentclass[twocolumn,aps,prc,showpacs,superscriptaddress,preprintnumbers,floatfix,nofootinbib]{revtex4}
\usepackage{epsfig,graphics}
\usepackage{graphicx}
\usepackage{dcolumn}
\usepackage{bm}
\usepackage{amsmath}
\usepackage{amssymb}
\usepackage[usenames]{color}
\usepackage{ulem} 
\usepackage[colorlinks,linkcolor=blue,urlcolor=blue,citecolor=blue]{hyperref}
\usepackage{makeidx}
\makeindex
\usepackage{CJK}

\begin{document}

\begin{CJK*} {UTF8} {gbsn}

\title{$\alpha$-clustering effect on flows of direct photons  in heavy-ion collisions}

\author{C. Z. Shi (施晨钟)}
\affiliation{Shanghai Institute of Applied Physics, Chinese Academy of Sciences, Shanghai 201800, China}
\affiliation{Key Laboratory of Nuclear Physics and Ion-beam Application (MOE), Institute of Modern Physics, Fudan University, Shanghai 200433, China}
\affiliation{University of the Chinese Academy of Sciences, Beijing 100080, China}

\author{Y. G. Ma (马余刚)}\thanks{Author to whom all correspondence should be addressed. Email: mayugang@fudan.edu.cn}
\affiliation{Key Laboratory of Nuclear Physics and Ion-beam Application (MOE), Institute of Modern Physics, Fudan University, Shanghai 200433, China}
\affiliation{Shanghai Institute of Applied Physics, Chinese Academy of Sciences, Shanghai 201800, China}

\date{ \today}

\begin{abstract}
In this work, we reconstruct the $\gamma$-photon energy spectrum which is in good agreement with the experimental data of  $^{86}$Kr + $^{12}$C 
at  $E/A$ = 44 MeV in the framework of our modified EQMD model.
The directed flow and elliptic flow of free protons and direct photons have been investigated by taking $\alpha$-clustering structure of $^{12}$C into account.
Comparing with the free proton, the direct photon flows give a clearer information about early stage of nuclear reaction.
Difference of collective flows between different configurations of $^{12}$C is observed in this work. This indicates that collective flows of direct photons are sensitive to the initial configuration, therefore
the $\gamma$ bremsstrahlung process might be taken as  an alternative probe to investigate $\alpha$-clustering structure in light nucleus from heavy ion collisions at Fermi-energy region.

\end{abstract}


\maketitle

\section{Introduction}
The $\alpha$-clustering configuration in light nucleus has attracted a lot of attention for a very long time. In 1930s, even before the discovery of neutron, this concept has been earliest assumed by Gamow \cite{Gamow1930} and discussed by Bethe and Bacher \cite{RMP.8.82, RMP.9.69} due to the highest stability of the $\alpha$-cluster around neighbouring light nuclei. Following this idea, a picture of nuclear molecular state \cite{alpha_alpha} based on the $\alpha$-$\alpha$ interaction which presents as a repulsive force in short and long ranges or as an attractive force in intermediate range have been established. Successfully, such a simple description gave a good agreement with binding energies of the namely saturated nuclei composed of multiple-$\alpha$ clusters. However, those $\alpha$-clustering structure has not been observed in ground states of nucleus, but observed in excited states which close to the $\alpha$-decay thresholds. Among the numerous $\alpha$-clustering nuclei, $^{12}$C is one of the most popular candidates. This is related to its important role in the process of nucleosynthesis process in astrophysics \cite{Hoyle,Jin,LiuHL1,LiuHL2}, i.e. the triple-$\alpha$ process.

Nowadays, there are many models or theories having capabilities to produce $\alpha$-clustering nuclei, such as  antisymmetrized molecular dynamics (AMD) \cite{AMD_PPNP}, fermionic molecular dynamics (FMD) \cite{FMD_RMP} and an extended quantum molecular dynamics (EQMD) \cite{EQMD} etc. Specifically for  $^{12}$C, the  chain structure is usually considered as an excited state,  so we just consider the triangle- and sphere-configuration in this work. Recently, a lot of studies aiming at $\alpha$-cluster have been done in a framework of transport models. In those works, giant dipole resonance \cite{He_wb_gdr_prl, HeWB2}, photon disintegration \cite{Huang_bs}, HBT correlation \cite{hejj,Huang2,Huang3} and collective flows \cite{PRL1,zhangsong_flow_prc, zhangsong_flow_epja,PRC2, GCC_PRC, Guo_2017,LiYA} were suggested as useful probes to investigate the $\alpha$-clustering configuration in light nucleus.

On the other hand, for anisotropic flow itself, also namely collective flow, it is a widely investigated quantity in nuclear physics from more than ten $A$MeV to several $A$ TeV \cite{Reisdorf, Herrmann}. Its earliest ideas \cite{1958Collective} can be traced back to the 1950s, when fluid dynamical models were used to describe the collision between nucleons and nuclei. In 1970s, this concept has been extended to heavy ion reactions \cite{Greiner_flow}, but it was not convincingly observed in the laboratory until 1990s \cite{evidence_flow_1, evidence_flow_2}. Nowadays, it is well known that the anisotropic flow can be taken as an important observable for nuclear physics which could study in-medium cross section \cite{ogilvie_balanceenergy}, viscosity coefficient \cite{Zhou}, equation of state (EOS) \cite{westfall} and reaction mechanism \cite{libaoan_flow} in the low- and medium-energy region. Moreover, such a collective movement is also significant for studying QGP at relativistic energy range due to its sensitivity to the early partonic dynamics \cite{Heinz,NST1,NST2}. Based on this reason, the idea of using anisotropic flow to investigate the different initial geometry of light nucleus \cite{PRL1, zhangsong_flow_prc, zhangsong_flow_epja} has been proposed at relativistic energies.

In such works, all observed particles are hadrons which would be influenced by the surround nuclear matter in final stage especially at Fermi energy region.
However, direct photons produced in the early stage of heavy ion reaction is seldom influenced by the surround nuclear matter. In our previous articles \cite{LiuGH2008, LIU2008312, MaYG2012, wangss_flow}, an anti-correlation between direct photon and free proton has been observed in symmetric collision system. It indicates that direct photons are highly related to the collective flow of nucleons, and can be taken as an alternative probe to find a trace of $\alpha$-cluster structure in light nuclei.

In the present work, the geometry effects of two different configurations on collective flow of direct photons for the reaction of $^{86}$Kr + $^{12}$C at $E/A$ = 44 MeV \cite{BERTHOLET1987541} are studied. The article is arranged as follows: the model and method is briefly introduced in Section \ref{sec:2}, the results and discussion are given in Section \ref{sec:result} and a summary is given in Section \ref{sec:con}.

\section{The model and method description}
\label{sec:2}

\subsection{EQMD model}
\label{EQMD}
The EQMD model \cite{EQMD} is one kind of the QMD-type models whose equations of motion are determined by the time-dependent variational principle (TDVP) \cite{FMD_RMP}. In the EQMD model, the nucleon is represented by a coherent state and the total system is a direct product of each nucleon which is similar to the most QMD-type models. To satisfy the ground state fermion properties, a phenomenological Pauli-potential has been added. Besides,  dynamic wave packets are implied in this model, replacing fixed wave packets in traditional QMD modes. The  nucleon wave function and  total wave function can be written as follows

\begin{equation}
\begin{aligned}
{{\rm{\varphi }}_i}\left( {{{\bf r}_i}} \right) = & {\left( {\frac{{{v_i} + v_i^ * }}{{2\pi }}} \right)^{3/4}}\exp \left[ { - \frac{{{v_i}}}{2}{{\left( {{{\bf r}_i} - {{\bf R}_i}} \right)}^2} + \frac{i}{\hbar }{{\bf P}_i} \cdot {{\bf r}_i}} \right],\\
\Psi=&\prod_i\varphi_i(\mathbf{r}_i).
\end{aligned}
\label{eq_eqmd}
\end{equation}
Here $\mathbf{R}_i$ and $\mathbf{P}_i$ are mean value of wave packet belong to the $i$-th nucleon in the phase space. Correspondingly, $\upsilon_i = \frac{1}{\lambda_i} + i\delta_i$ is its complex wave packet width. Under the time dependent variation principle, the propagation of each nucleon can be described as follows:

\begin{equation}
\begin{aligned}
\dot{\mathbf{R}}_{i} = \frac{\partial H}{\partial \mathbf{P}_{i}}+\mu_{\mathrm{R}} \frac{\partial H}{\partial \mathbf{R}_{i}}, ~~~& \dot{\mathbf{P}}_{i}=-\frac{\partial H}{\partial \mathbf{R}_{i}}+\mu_{\mathrm{P}} \frac{\partial H}{\partial \mathbf{P}_{i}} \\
\frac{3 \hbar}{4} \dot{\lambda}_{i} = -\frac{\partial H}{\partial \delta_{i}}+\mu_{\lambda} \frac{\partial H}{\partial \lambda_{i}}, ~~~& \frac{3 \hbar}{4} \dot{\delta}_{i}=\frac{\partial H}{\partial \lambda_{i}}+\mu_{\delta} \frac{\partial H}{\partial \delta_{i}}.
\end{aligned}
\label{eq_motion}
\end{equation}
Here $H$ is the expected value of Hamiltonian, $\mu_{\mathbf{R}}$, $\mu_{\mathbf{P}}$, $\mu_{\lambda}$ and $\mu_{\delta}$ are friction coefficients. In the friction cooling process, these coefficients are negative for getting a stable nucleus, while in the followed nuclear reaction simulation stage, these coefficients are zero value to keep energy  conservation of system. The expected value of Hamiltonian of the EQMD model can be written as follow:

\begin{equation}
\begin{aligned}
H & = \left\langle\Psi\left|\sum_{i}-\frac{\hbar^{2}}{2 m} \nabla_{i}^{2}-\hat{T}_{\mathrm{zero}}+\hat{H}_{\mathrm{int}}\right| \Psi\right\rangle \\
& =\sum_{i}\frac{\mathbf{P}_{i}^{2}}{2 m}+\frac{3 \hbar^{2}\left(1+\lambda_{i}^{2} \delta_{i}^{2}\right)}{4 m \lambda_{i}}-T_{\mathrm{zero}}+H_{\mathrm{int}},
\end{aligned}
\label{eq_h}
\end{equation}
where the first three terms represent the expected value of kinetic energy of each nucleon.
The first term is the center momentum of wave packet $\langle \hat{\mathbf{p}}_i^2\rangle/2m$, the second term is the contribution of dynamic wave packet $(\langle \hat{\mathbf{p}_i}^2\rangle-\langle\hat{\mathbf{p}_i}\rangle^2)/2m$, and the third term $-T_\mathrm{zero}$ is the zero-point center-of-mass kinetic energy, its detail form can be written as
\begin{equation}
\begin{aligned}
T_{\mathrm{zero}}&=\sum_{i} \frac{t_{i}}{M_{i}}\\
t_{i}&=\frac{\left\langle\phi_{i}\left|\hat{\mathbf{p}^2}\right| \phi_{i}\right\rangle}{2 m}-\frac{\left\langle\phi_{i}|\hat{\mathbf{p}}| \phi_{i}\right\rangle^{2}}{2 m}.
\end{aligned}
\label{eq:zero}
\end{equation}
Here $M_i$ is the ``mass number" and its detailed definition as follows \cite{EQMD}

\begin{equation}
\begin{aligned}
M_{i}&=\sum_{j} F_{i j} \\
F_{i j}&=\left\{\begin{array}{ll}
1 & \left(\left|\mathbf{R}_{i}-\mathbf{R}_{j}\right|<a\right) \\
e^{-\left(\left|\mathbf{R}_{i}-\mathbf{R}_{j}\right|-a\right)^{2} / b} & \left(\left|\mathbf{R}_{i}-\mathbf{R}_{j}\right| \geqslant a\right),
\end{array}\right.
\end{aligned}
\end{equation}
where the parameters $a=1.7$ fm and $b=4\ \mathrm{fm}^2$.

\begin{figure}[htbp]
\resizebox{8.6cm}{!}{\includegraphics{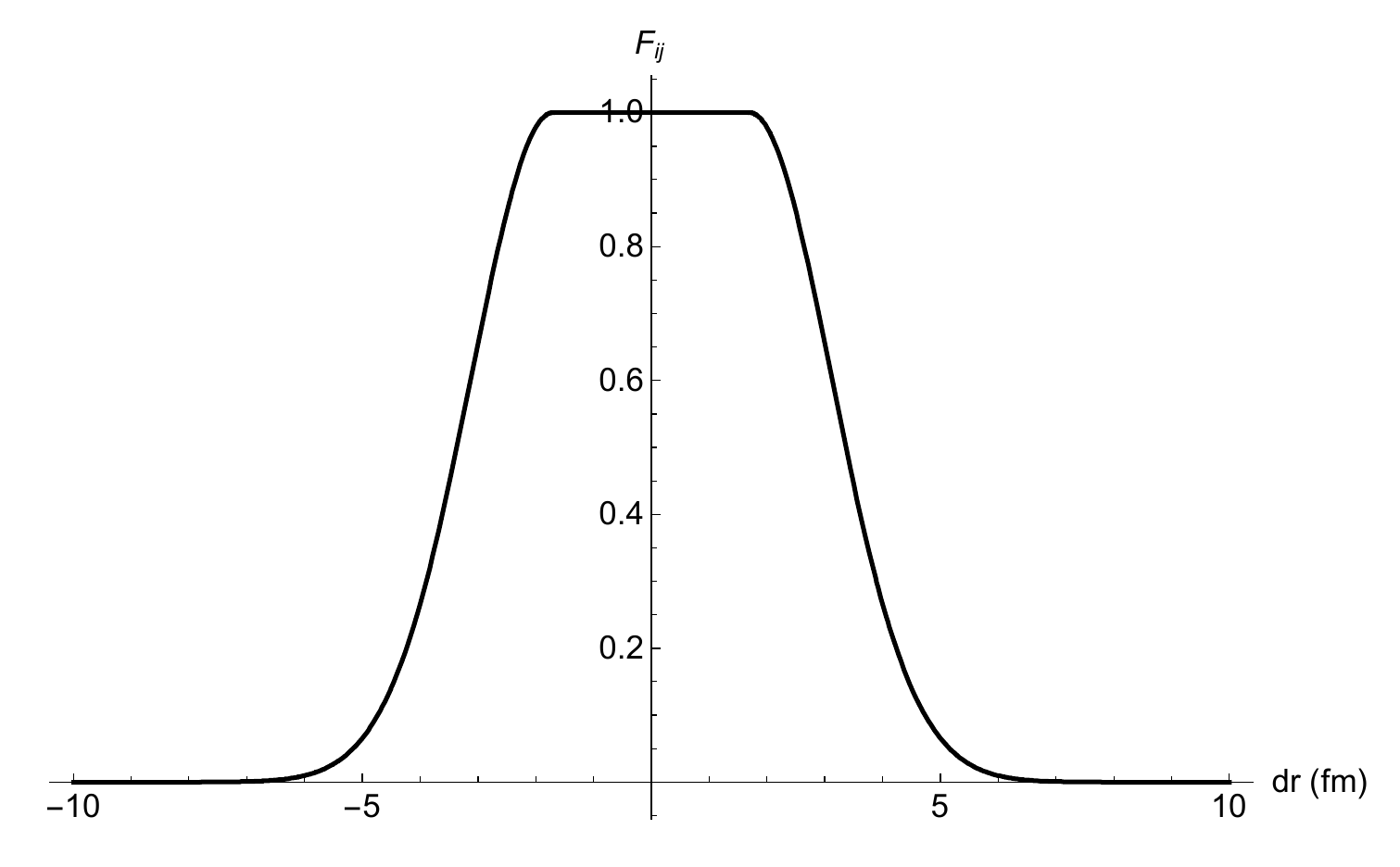}}
\caption{The ``mass" number as a function of the distance between two nucleons. Here $dr=R_i-R_j$.}
\label{zero_energy}
\end{figure}

In addition, we also show the "mass number" as a function of the distance between two nucleons. It is obviously that the "mass number" increases as the two nucleons approach; Otherwise, it decreases. Comparing the second term with the third one, it is not hard to find that they are only different with a factor of $\frac{1}{M_i}$ and a minus sign. When $i$ and $j$ are the same particle, its value gets 1. So, if a nucleus break away from the collision and becomes as a free particle, the "mass number" drops to 1. Then, the zero-point kinetic energy will offset the second item contribution. This is related to our momentum re-extraction later.

The last term represents the mean-field potential adopted in the EQMD model. In this model, only a very simple potential form was used, e.g. two-body and density dependent terms from Skyrme, Coulomb, symmetry and Pauli potential. The detailed description can be found in Ref.~\cite{EQMD}. We use the second set of parameter in this work.
It should be emphasized that the dynamical wave packets also have contributions on the kinetic energy. This treatment is very different from the most QMD-type models. This is the reason why the energy spectrum of direct photon is unreasonable which reported in our previous work \cite{shi_wavepacket}. Based on this reason, it is necessary to consider this effect in in-elastic process self-consistently. As to why the zero-point kinetic energy introduced, please refer to the article \cite{Ono:1992uy}.

\subsection{Bremsstrahlung process in the EQMD}
\label{bremsstrahlung_process}

Only the first instant proton-neutron collision has been considered in this paper to make bremsstrahlung  contributions for direct photons. The elementary double differential cross section we adopted was improved by Bauer {\it et al.} \cite{Bauer} as
\begin{equation}
\frac{d^{2} \sigma^{\text {elem }}}{d E_{\gamma} d \Omega_{\gamma}}=\alpha_c \frac{R^{2}}{12 \pi} \frac{1}{E_{\gamma}}\left(2 \mathbf{\beta}_{f}^{2}+3 \sin ^{2} \theta_{\gamma} \mathbf{\beta}_{i}^{2}\right).
\label{eq_element}
\end{equation}
Here $R$ represents the radius of the hard sphere, $\alpha_c$ represents the fine structure constant, $\beta_{i}$ and $\beta_{f}$ represent the velocity of proton at initial and final stage in the rest proton-neutron frame. $\theta_{\gamma}$ represents the angle between the outgoing photon and the direction of the incident proton in the same rest frame. Then, combining with the Pauli-blocking in finial state of scattered proton and neutron, the probability for a photon emitted from a $p$-$n$ collision can be deduced as follow:
\begin{equation}
\begin{aligned}
\int & \frac{d \Omega_{e}}{4 \pi} \frac{1}{\sigma_{N N}} \frac{d^{2} \sigma^{\text {elem }}}{d E_{\gamma} d \Omega_{\gamma}} \\
& \times\left[1-S_{3}\left(\mathbf{r}_3, \mathbf{k}_{3}, t\right)\right]\left[1-S_{4}\left(\mathbf{r}_4, \mathbf{k}_{4}, t\right)\right].
\end{aligned}
\label{eq_probibility}
\end{equation}
Here $\sigma_{NN}$ is the elemental nucleon-nucleon cross section, $\mathbf{r}_3$, $\mathbf{r}_4$, $\mathbf{k}_3$ and $\mathbf{k}_4$ are scattered nucleons' phase-space coordinates in final state and $S_3$, $S_4$ are their effective occupation fraction, respectively. The parameters we adopted are the same as the choice of Bauer {\it et al.} \cite{Bauer}. From our previous work \cite{shi_wavepacket}, the general approach treating two-body inelastic scattering is not applicable to the EQMD model due to the effects of dynamic wave packets. Based on this reason, a feasible method and the proof of its effectiveness have been offered. In order to avoid considering  of geometric condition of two-body scattering after sampling, a slightly adjustment has been made in this work. We assume that the two particles are in the same position after sampling if  they are  judged to have collision. In other words, a position sampling of scattering nucleons has been done before their momentum sampling. Then, the coupling term of coordinate and momentum which has been neglected before has been introduced. The Wigner function of the EQMD can be written as \cite{shi_wavepacket}
\begin{equation}
\begin{aligned}
\mathrm{f}(\mathbf{r}, \mathbf{p})&=\left(\frac{1}{\pi \hbar}\right)^{3} \sum_{i}^{A} \exp [ -\frac{1+\lambda_{i}^{2} \delta_{i}^{2}}{\lambda_{i}}\left(\mathbf{r}-\mathbf{R}_{i}\right )^{2}\\
&-\frac{2 \lambda_{i} \delta_{i}}{\hbar}\left(\mathbf{r}-\mathbf{R}_{i}\right)\left(\mathbf{p}-
\mathbf{P}_{i}\right) -\frac{\lambda_{i}}{\hbar^{2}}\left(\mathbf{p}-\mathbf{P}_{i}\right)^{2}].
\end{aligned}
\label{eq_wigner}
\end{equation}
Here $\mathrm{f}$ represent the phase-space density which is the closest analogue to classical phase-space density. The position of the scattered nucleons can be decided by density overlap of two nucleons as follows
\begin{equation}
\begin{aligned}
&f_{i, j}(\mathbf{r}) =\frac{\rho_{i}(\mathbf{r}) \rho_{j}(\mathbf{r})}{\int \rho_{i}(\mathbf{r}) \rho_{j}(\mathbf{r}) d \mathbf{r}}\\
&=\left(\frac{\lambda_{i}+\lambda_{j}}{\pi \lambda_{i} \lambda_{j}}\right)^{3/2} \exp \left\{-\frac{\lambda_{i}+\lambda_{j}}{\lambda_{i} \lambda_{j}}\left(\mathbf{\mathbf{r}}-\frac{\lambda_{i} \mathbf{R}_{j}+\lambda_{j} \mathbf{R}_{i}}{\lambda_{i}+\lambda_{j}}\right)^{2}\right\}.
\end{aligned}
\label{eq:sample_r}
\end{equation}
Here $\rho_{i}(\mathbf{r})\rho_{j}(\mathbf{r})$ is the density overlap, and the integration on the denominator is to ensure probability normalization. Then, it is easy to deduce the momentum distribution of the $i$-th nucleon belongs the conditional probability rule,
\begin{equation}
\begin{aligned}
&f_{i}(\mathbf{p}) =\frac{w_{i}(\mathbf{r},\mathbf{p})}{w_{i}(\mathbf{r})} \\
&=\left(\frac{\lambda_{i}}{\pi \hbar^{2}}\right)^{3 / 2} \exp \left\{-\frac{\lambda_{i}}{\hbar^{2}}\left[p-\left(\mathbf{P}_{i}-\delta_{i} \hbar \mathbf{r}+\delta_{i} \hbar \mathbf{R}_{i}\right)\right]\right\},
\end{aligned}
\label{eq:sample_p}
\end{equation}
where $w_{i}(\mathbf{r},\mathbf{p})$ represents the Wigner density of the $i$-th nucleon at point $(\mathbf{r},\mathbf{p})$ in phase space. Comparing with previous work, it needs one more sampling in coordinate space. $f_i(\mathbf{p})$ represents the momentum distribution of the $i$-th nucleon with coordinate position at $\mathbf{r}$. Now we get the momentum of the $i$-th nucleon randomly as,
\begin{equation}
\begin{aligned}
\mathbf{r}_{i} =& \mathbf{r} \\
\Delta\mathbf{p}=& \mathbf{p}-\mathbf{P}_i\\
\mathbf{p}_{i} =& \mathbf{P}_{i}+\Delta \mathbf{p} \times \sqrt{1-\frac{1}{M_i}}.
\end{aligned}
\label{eq:9}
\end{equation}
The square root term is taking into account zero-point center-of-mass kinetic energy Eq.~\ref{eq:zero}.
This treatment can be both applicable to free nucleons and the nucleons in medium. Besides, when a nucleon becomes free, it will automatically degenerate to the traditional form, i.e., using the center momentum of wave packet as a nucleon's momentum, since its zero-point kinetic energy covers its dynamic wave packet's contribution on kinetic energy, see subsection \ref{EQMD}. This property is not intended by us. It is automatic satisfaction when we consider the zero-point kinetic energy which is respect to the design of Hamiltonian by Maruyama {\it et al.} \cite{EQMD}. When a nucleon becomes a free nucleon, we can directly use its central coordinates to calculate its collective motion, which is consistent with  works of Guo {\it et al.} \cite{GCC_PRC, Guo_2017}.

Strictly speaking, the coordinate and momentum of a nucleon should be sampled adequately according to it own Wigner function (Eq.~\ref{eq_wigner}) like most BUU-type models do. Meanwhile, it should take into account the zero-point kinetic energy also. We show the deduction in Appendix \ref{appendix}. However, for simplify the calculation, we only sample once following Eqs. \ref{eq:sample_r}, \ref{eq:sample_p} and \ref{eq:9} which sampling of coordinate space is limited. Another advantage for the Eq.~\ref{eq:sample_r} is that the in-medium effect can be easy to be included by assuming those two nucleons collide at the same point.

Besides, the Pauli-blocking needs to be reconsidered because of the information loss of the wave packet after sampling. In this work, we calculate the effective occupation fraction as follows:
\begin{equation}
S_i = h^3 \times\sum_{j \ne i}\delta_{\tau_{i},\tau_{j}} \delta_{s_i,s_j} w_j(\mathbf{R}'_i,\mathbf{P}'_i).
\end{equation}
Here $\mathbf{R}'_i$ and $\mathbf{P}'_i$ is the coordinate and momentum of the scattered nucleons in the final state. $w_j$ is the density of Wigner function contributed by the other nucleons to the phase-space point $(\mathbf{R}'_i,\mathbf{P}'_i)$. $h^3$ is a volume size. A more detailed description can be found in our previous paper \cite{shi_wavepacket}. However, it should be admitted that this method is a very simple way to include fermonic property of nucleons.

\subsection{Anisotropic flow of direct photons}

\label{flow_photon}
Anisotropic flow can be characterized by the Fourier expansion of the emitted particles as follows \cite{PhysRevC.58.1671}
\begin{equation}
E \frac{d^{3} N}{d^{3} p}=\frac{1}{2 \pi} \frac{d^{2} N}{p_{T} d p_{T} d y}\left(1+\sum_{i=1}^{N} 2 v_{n} \cos \left[n\left(\phi-\Psi_{n}\right)\right]\right),
\end{equation}
here $\phi$ is the azimuthal angle between the emitted particle and reaction plane, and $v_n$ represents different orders of the Fourier coefficients which can be deduced if assuming the reaction plane is the  $x-z$ plane, i.e. $\Psi_{n}=0$ in our calculation, as follows
\begin{equation}
\begin{aligned}
&v_{1} = \langle\cos([\phi-\Psi_1])\rangle = \langle\frac{p_{x}}{p_{T}}\rangle \\
&v_{2} = \langle\cos(2[\phi-\Psi_2])\rangle = \langle\frac{p_{x}^{2}-p_{y}^{2}}{p_{T}^{2}}\rangle .
\end{aligned}
\end{equation}
Here we only show the first- and second-order coefficients of the Fourier expansion, $\langle...\rangle$ represents statistical averaging over all emitted particles in all events and $p_T = \sqrt{p_x^2+p_y^2}$ is the transverse momentum of emitting particles.
By convention, it defines directed flow of the particles emitted forward as a positive value when the repulsion dominates for free proton or neutron \cite{Reisdorf}. While a negative value is presented when the attraction dominates for symmetry system. We are also following this rule in this paper although an asymmetry system is used. The actual situation of hadrons is more complicated for an asymmetry system.

In Ref.~\cite{PRL1}, they use a carbon to collide lead in relativistic energy region. Then, the shape of the created fireball in the transverse plane reflects the shape of $^{12}$C. The formed hot zone expands outward along its density gradient direction. Therefore, using a harmonic flow analysis can indirectly reflect the spatial deformation of the initial state \cite{PRL1}. However, in the low energy region where the attraction dominates, the formation mechanism of collective flow is very different \cite{Ma1,Ma2}. This is the result of competition between the mean-field potential and two-body collision. The created compressed zone rotates along $y$ direction perpendicular to the reaction plane due to the angular momentum effect by a finite impact parameter. Finally, the particles prefer  side splashing in reaction plane.

\section{Results and Discussion}
\label{sec:result}

\begin{figure}[htbp]
\resizebox{8.6cm}{!}{\includegraphics{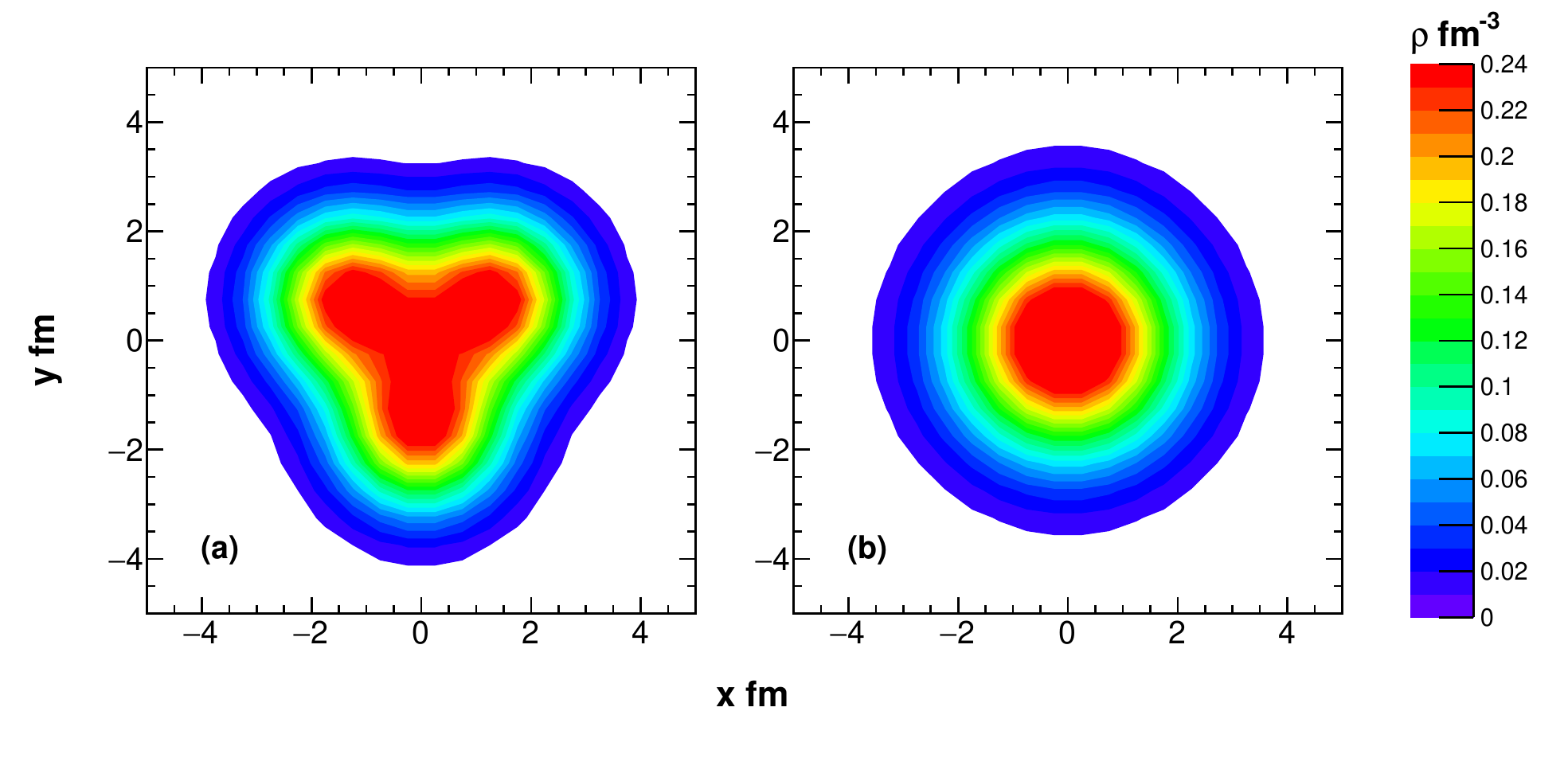}}
 \caption{The density distribution of $^{12}$C with configurations of $\alpha$-clustered triangle (a) and of sphere (b). The configuration of $\alpha$-clustered $^{12}$C is rotated to a plane which is perpendicular to its symmetry axis in coordinate space.}
\label{configuration}
\end{figure}

Figure \ref{configuration} shows the density contour of different configurations of $^{12}$C where the panel (a) corresponds to triangle configuration and the panel (b) to the sphere configuration, respectively. Obviously, there are three cores far away from the central region in triangle configuration of $3\alpha$ clusters. On the contrary, there is only one core in the central region in the sphere configuration case. Those two configurations also appear in the AMD simulation \cite{furuta_monopole_2010}. With the increase of the excitation energy, the transition from the sphere configuration to the triangle configuration has been observed. However, it worths noting that the ground state of $^{12}$C initialized by the friction cooling process in the EQMD model is triangle configuration of $3\alpha$. Strictly speaking, the EQMD model does not provide the complete Fermi properties due to the phenomenological Pauli-potential used instead of the antisymmetry of wave function. But in comparison with the AMD and the FMD, it has more significant computing advantages.

The different kinds of configuration given by the EQMD model is randomly. Before the friction cooling process,  nucleons' position is sampled within a solid sphere, then  their momentum is sampled according to the Thomas-Fermi gas model. The cooling path will be influenced by the initial coordinate and momentum. Therefore all configurations could been created during a batch initialization processes, but just the specific configurations are chosen as an input initial nucleus.

\begin{table}[htbp]
\caption{ The values of binding energy and r.m.s radii of different configurations initialized by the EQMD as well as the experimental measurement. }
\label{tab:1}
\begin{tabular}{|c|c|c|}\hline
& $E/A$ (MeV) & $R.M.S$ (fm) \\\hline
triangle &-7.721 &2.313  \\\hline
sphere &-6.982 &2.333 \\\hline
Exp. &-7.68 &2.46 \\\hline
\end{tabular}
\end{table}

Table \ref{tab:1} lists the binding energy and the root mean square (r.m.s) radius of $^{12}$C from experimental measurement as well as  the results calculated by the EQMD for different configurations. The r.m.s. radii of different configurations are near. Although the binding energy of sphere configuration adopted in this work is about 0.8 MeV higher than that of triangular configuration, they are both close to the experimental value. It should be emphasized that it does not study the energy level of different configurations exactly, but focus on the geometric effects during the heavy ion collision in this paper.

\begin{figure}[htbp]
\resizebox{8.6cm}{!}{\includegraphics{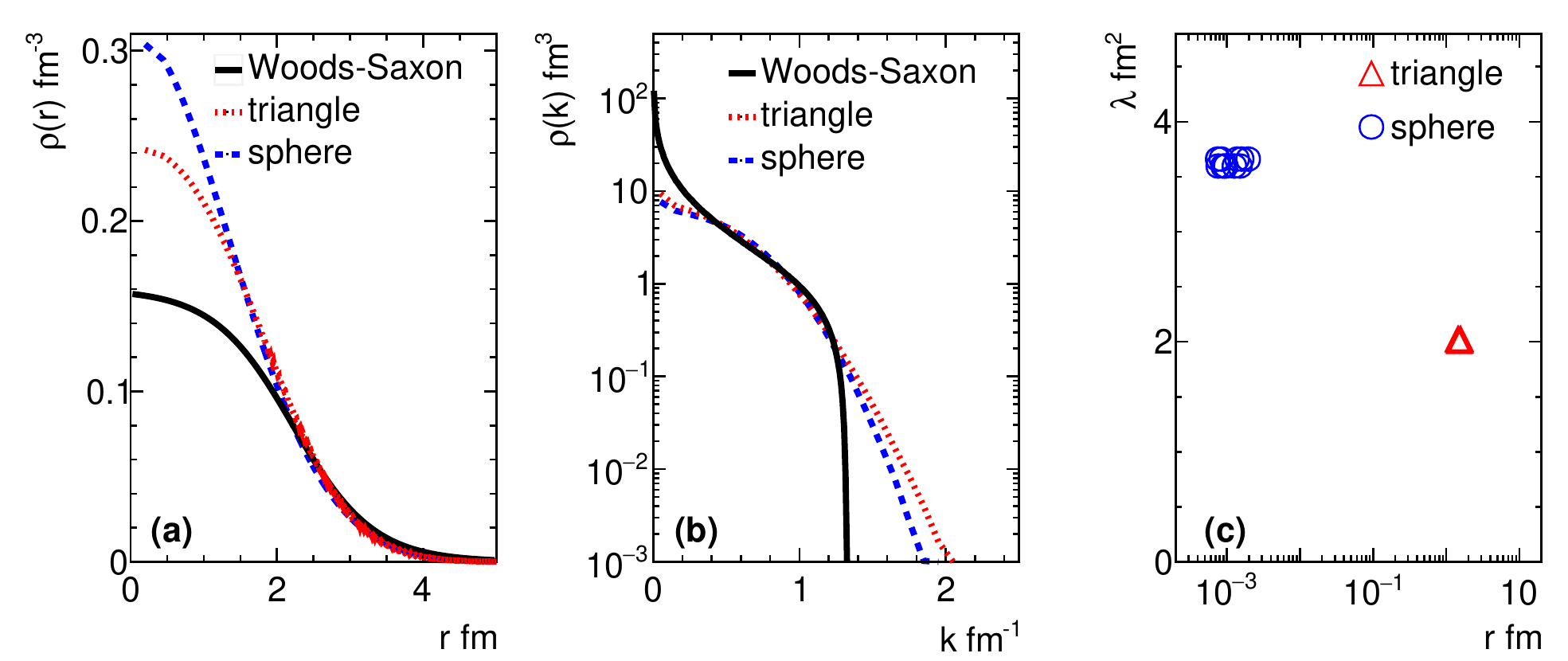}}
 \caption{The density (a), momentum (b) and width of wave packets (c) distribution of different configurations of $^{12}$C. It does not distinguish proton and neutron in this paper. The solid black line shown in panel (a) and (b) represent the density and momentum distribution calculated by the Woods-Saxon distribution and under the free Fermi-gas approximation.}
\label{distribution}
\end{figure}

Figure \ref{distribution} shows the density (a), momentum (b) and the width of wave packets (c) distribution of different configurations for $^{12}C$. For comparison, the Woods-Saxon situation has also been drawn as a solid black line in panel (a) and (b). In panel (a), the differences of density distribution among three configurations are  clearly observed. In the case of the Woods-Saxon distribution, the density keeps almost constant  of $0.16/\mathrm{fm}^3$ near central region. However, in the case of triangular configuration, the density is higher. And in the case of spherical configuration, the highest local density in center of nucleus is reached. It indicates that the nucleons are closer together in a spherical configuration than them in the other situations. In panel (b), the momentum distribution of three configurations are shown. Comparing with the Woods-Saxon case, the momentum distribution initialized in the EQMD has a larger upper bound. Besides, this upper bound is highest in the triangle case related to the width distribution of wave packets shown in the panel (c). In the triangle case, the width of each particle is about $2\mathrm{fm}^{2}$, while it is about $3.6\mathrm{fm}^{2}$ in the sphere case. A large wave packet width leads to a narrower momentum distribution due to the uncertainty principle. On the other hand, although the spherical nucleus is compressed more tightly, the wave packet is also wider, so the r.m.s radii are  similar for those two different configurations.

\begin{figure}[htbp]
\resizebox{8.6cm}{!}{\includegraphics{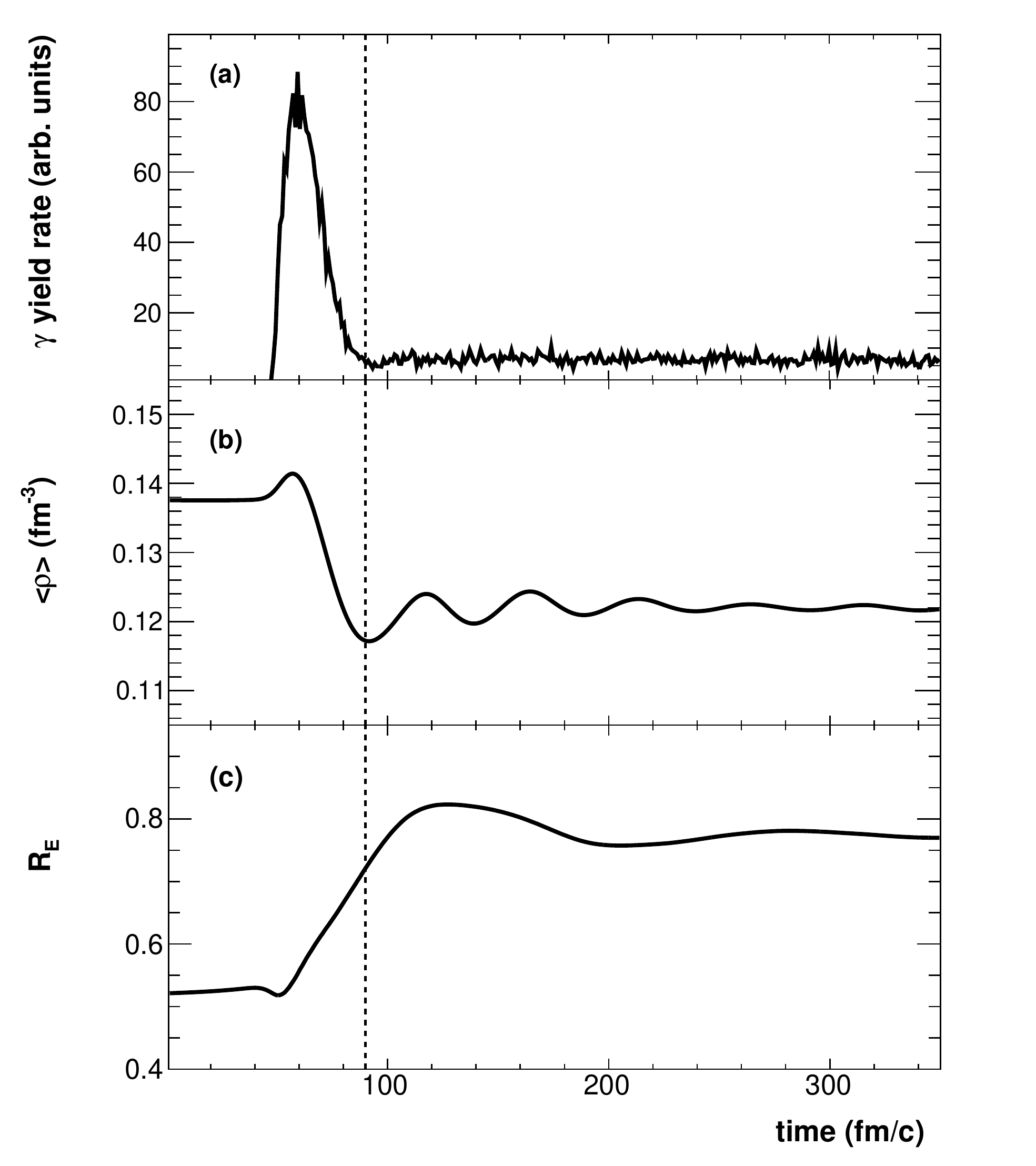}}
\caption{The time evolution of $\gamma$-production rate (a), average density (b) and stopping power (c) for  $^{86}$Kr+$^{12}$C at $E/A$ = 44 MeV. The black dot line near time $t\approx90 fm/c$ represents the end of the first compression state.}
\label{yield}
\end{figure}

The time evolution of the hard photon production rate (a) for $^{86}$Kr + $^{12}$C at $E/A$ = 44 MeV, the average density (b) and the stopping power (c) for this system are shown in Figure \ref{yield}. We calculate the average density as follow:
\begin{equation}
\overline{\rho}=\frac{\int \rho^{2} d \mathbf{r}}{\int \rho d \mathbf{r}}=\frac{\sum_{i, j}^{A} \int \rho_{j}(\mathbf{r}) \rho_{i}(\mathbf{r}) d \mathbf{r}}{A}.
\end{equation}
Here $A$ is the size of reaction system, i.e., the number of total nucleons in system.
It is clean that there is a strong positive correlation between $\gamma$ yield and the compression density which is owing to the frequent $p-n$ collision. The minimum value of $\gamma$ yield rate and the average density of system near $t$ = 90 fm/c where the black dot line locates at. It means that the separation of the target-like and project-like nucleus happens. In other words, it constrains a time selection range of photon. We only select those photons produced during the first compression stage of the system for analysis. At the same time, what's interesting is that the stopping power has not reached the maximum value during this process. In literature, there are several forms of stopping power definition. In this work, we adopt the energy-based isotropic ratio form \cite{zhanggq_2011} written as
\begin{equation}
R_{E} = \frac{\sum E_{T}^{i}}{2 \sum E_{l}^{i}}.
\end{equation}
Here $E_{T}^{i}$ represents the transverse kinetic energy which is perpendicular to beam direction, and $E_{l}^{i}$ represents the longitudinal kinetic energy which is parallel to the beam direction.
The stopping parameter is such a physical quantity which not only describes the stopping power of nuclear matter, but also indicates the system thermalization. One can expect $R_E = 1$ for complete thermal equilibration  and $R_E<1$ for partial equilibration.
In our case, it is obviously that the system has not yet achieved a complete equilibrium during the pre-equilibrium stage.

\begin{figure}[htbp]
\resizebox{8.6cm}{!}{\includegraphics{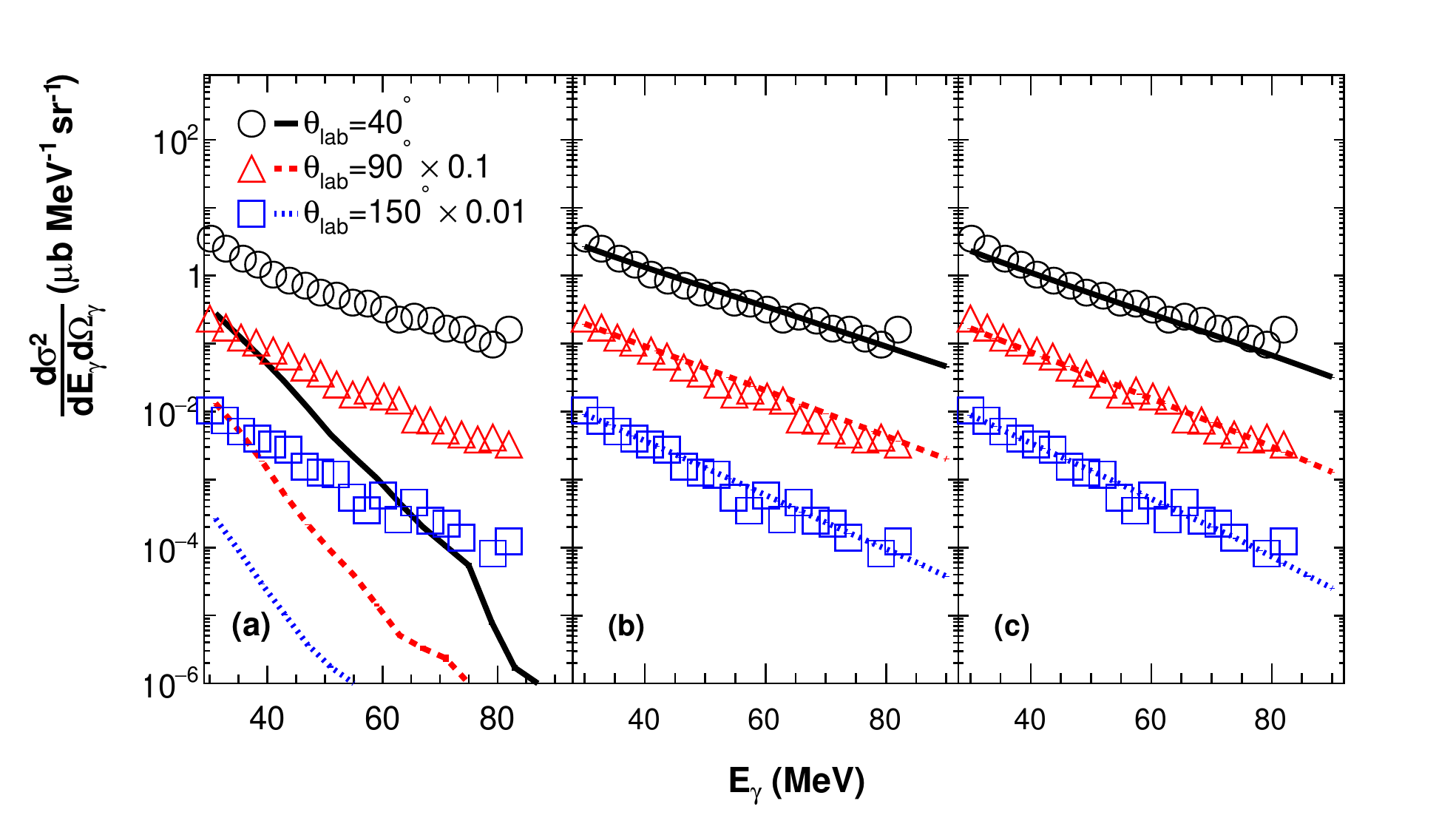}}
\caption{The energy spectrum of hard photons at angle $\theta_{\gamma} = 40^{\circ}, 90^{\circ}, 150^{\circ}$ for $^{86}$Kr + $^{12}$C at $E/A$ = 44 MeV. The results simulated by the EQMD without considering dynamical wave packet effects is shown in panel (a). The case of triangle configuration of $3\alpha$ with our modified EQMD is shown in panel (b), and the sphere configuration is shown in panel (c). The lines represent the results of theoretical calculation and the open markers represent the experimental data  taken from article \cite{BERTHOLET1987541} .}
\label{44_3}
\end{figure}

\begin{figure}[htbp]
\resizebox{8.6cm}{!}{\includegraphics{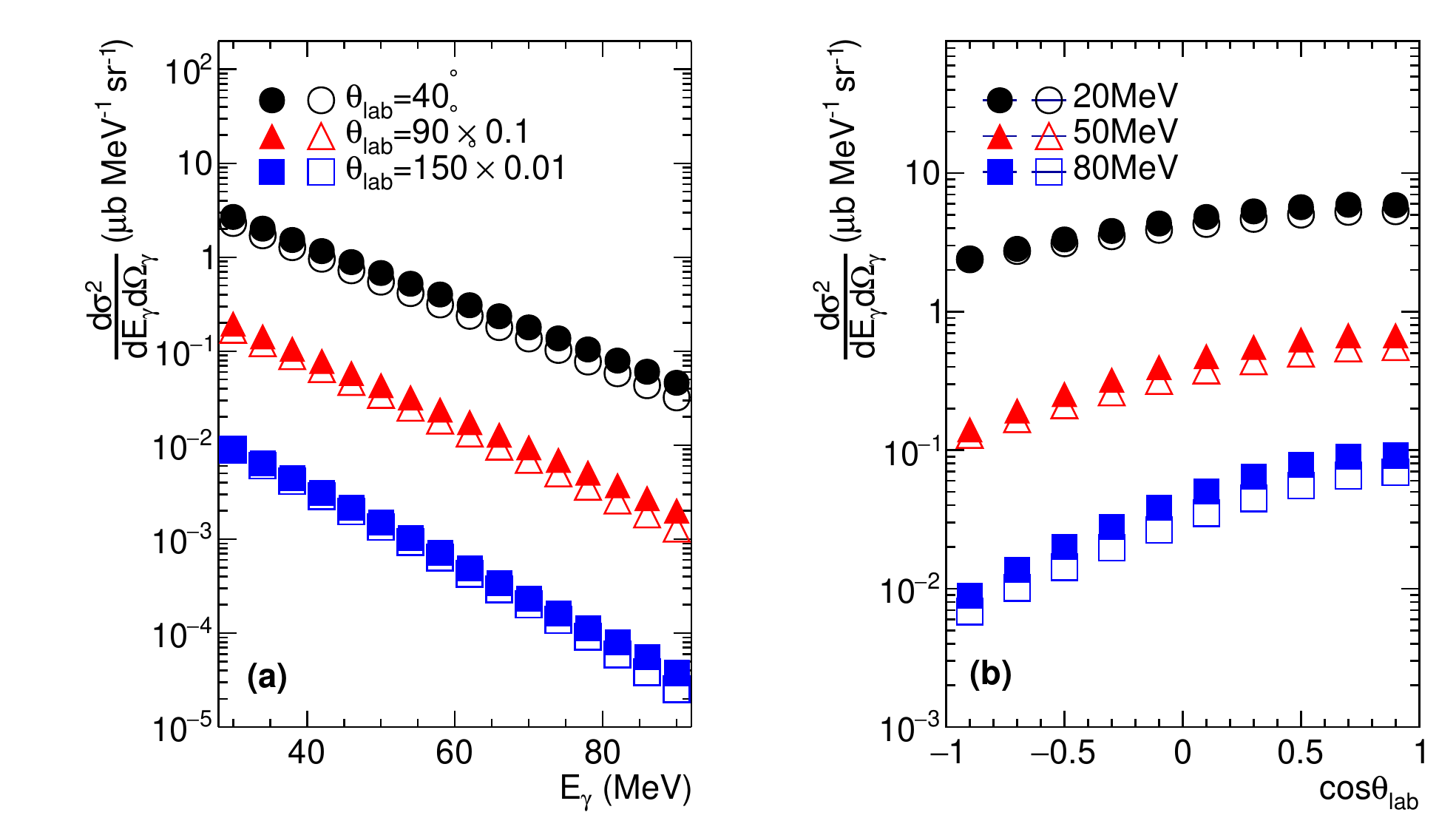}}
\caption{The energy spectrum (a) and laboratory angular distributions (b) of direct photons from the same reaction as Fig. \ref{44_3} for different initial configurations of $^{12}$C. The solid and open markers represent triangle and sphere configurations, respectively.}
\label{44_2}
\end{figure}

Figure \ref{44_3} shows the energy spectrum of hard photons emitted at angle $\theta_{lab} = 40^{\circ}, 90^{\circ}, 150^{\circ} $ in the Lab. frame for $^{86}$Kr + $^{12}$C at $E/A$ = 44 MeV. The results calculated without considering dynamical wave packet effects, i.e., only including the center momentum of wave packets, is plotted in panel (a),  and the same quantity simulated by the modified EQMD model with triangle  and sphere configurations are shown in panel (b) and (c), respectively. It is cleanly seen that  the photon yield is seriously underestimated in the first situation. However, the energy spectra are reconstructed reasonably and reproduced  experimental data no matter what initial configurations are adopted with our method described above. It confirms that the general method adopted by many QMD-type transport models which treats nucleon as a point rather than a packet in two-body scattering process would underestimate the available energy of in-elastic process in the framework of EQMD model which has been described in our previous work in more details \cite{shi_wavepacket}. It was noted that there is no contribution of wave packets to the kinetic energy for most QMD-type model. But the contribution of wave packet must be considered because the total kinetic energy is composed of three parts (Eq~.\ref{eq_h}).

Figure \ref{44_2} (a) shows energy spectrum comparison between triangle and sphere configuration, and the laboratory angular distributions (b) for the same reaction with photon energies of $E_\gamma$ = $20\pm3$ (cycle), $50\pm3$ (triangle), and $80\pm3$ (square) MeV is shown in panel (b). The solid and open markers represent triangle and sphere configuration  of $^{12}$C. It seems that
the energy spectrum of triangle configuration is slightly higher and harder than the case of sphere configuration. The difference appears gradually with the increase of $\gamma$ energy. And the same phenomenon is also visible in the angular spectrum in the Lab frame. This is closely related to figure \ref{distribution} (b) above, the upper bound of kinetic energy in triangular configuration is higher than it in the case of sphere configuration. So the nucleon can take a higher available energy in the case of triangle configuration.

\begin{figure}[htbp]
\resizebox{8.6cm}{!}{\includegraphics{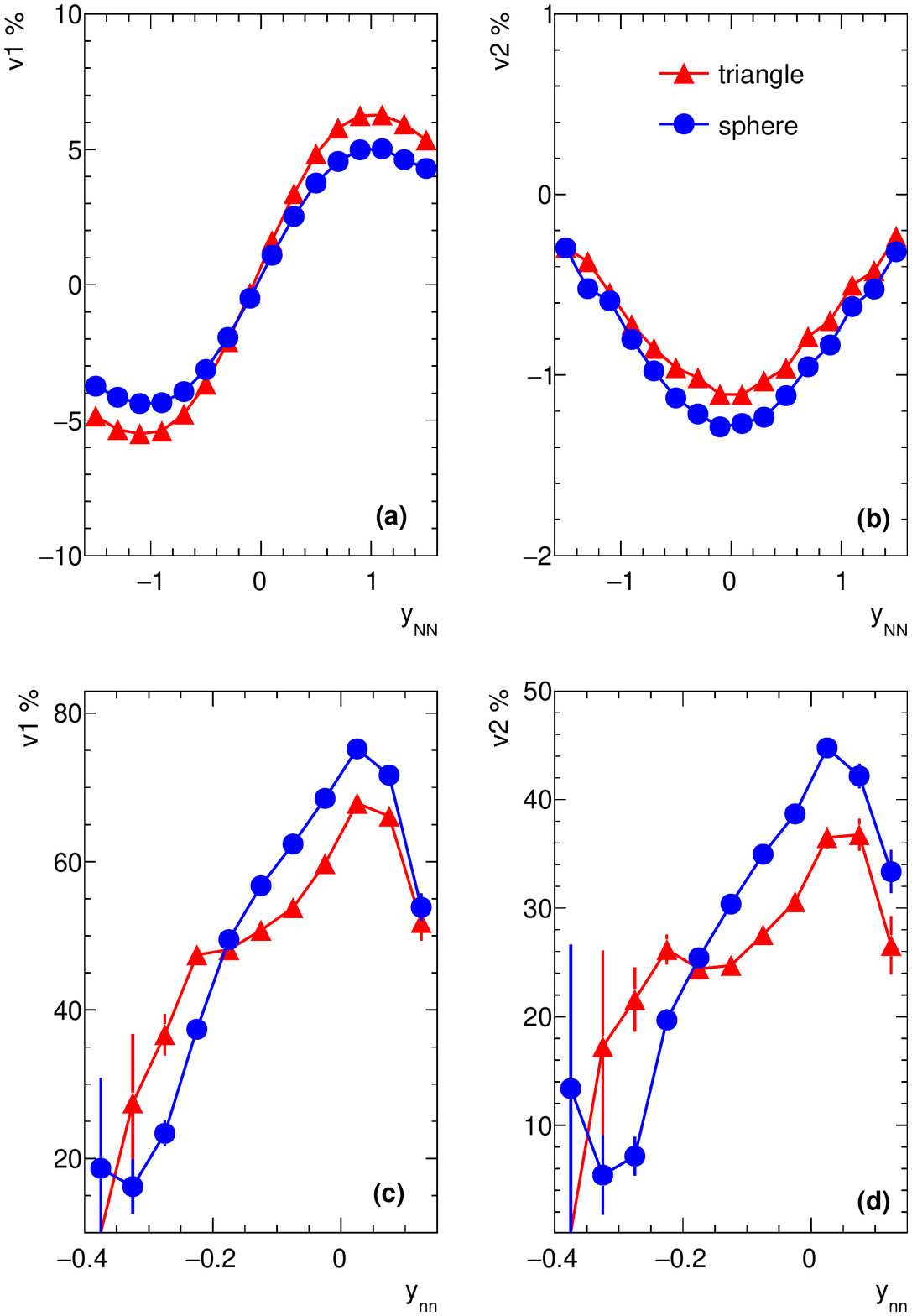}}
\caption{The directed (a) and elliptic flows (b) of direct photons versus rapidity for  $^{86}$Kr + $^{12}$C at  $E/A$ = 44 MeV and  impact parameter $b$ = 5.0 fm, respectively. The same observations of free protons in panel (c) and (d). Red triangle and blue sphere  markers represent the triangle and sphere configurations of $^{12}C$, respectively.}
\label{flow}
\end{figure}

Figure \ref{flow} (a) and (b) shows directed flow and elliptic flow of direct photons, respectively, and  (c) and (d) for  free protons. In figures, red triangle and blue sphere  markers represent the triangle and sphere configurations of $^{12}C$, respectively, and   $y_{NN}$ is the rapidity of direct photons in nucleon-nucleon center of mass system and $y_{nn}$ is the rapidity of emitted protons in nucleus-nucleus center of mass system. According to previous research results \cite{Guo_2017}, a large impact parameter $b$ = 5.0 fm is adopted. It is obviously that $v_1$ of direct photons has a clear $S$-shape curve even in the asymmetry system. However, the shape of the free protons' $v_1$ is irregular due to the asymmetry between projectile and target. This feature of direct photons can be explained by the geometrical equal-participant model \cite{nifenecker_equal} which describes a scenario that there are the same number of particles from projectile or target taking part in the bremsstrahlung processing. Our results confirm this theory very well, although there are different opinions have been proposed on light systems \cite{Gan1994}. Another important reason is that the direct photons are seldom absorbed by the surround nuclear matter, so the $v_1$ of direct photons can keep its shape. From the panel (a), a positive flow parameter of direct photons, i.e.  $F \equiv dv_{1}/dy_{NN}|_{y_{NN}=0}>0$, has been observed. However, it is hard to confirm the signal of free proton in anti-symmetry case (c) due to a seriously affected by surrounding matter. As well known, nuclear attraction dominates in the Fermi-energy region where there exists a competition between the mean field and nucleon-nucleon collision. General speaking, for a symmetric systems, a negative flow parameter is expected if attraction dominates. But in the case of an asymmetry system, this relationship seems unsatisfied again based on the EQMD simulation. Besides,  $v_1$ difference of direct photons between triangle and sphere configurations is observed in panel (a). The amplitude of directed flow calculated with the triangle configuration is slightly larger than that with sphere configuration.

\begin{figure}[htbp]
\resizebox{8.6cm}{!}{\includegraphics{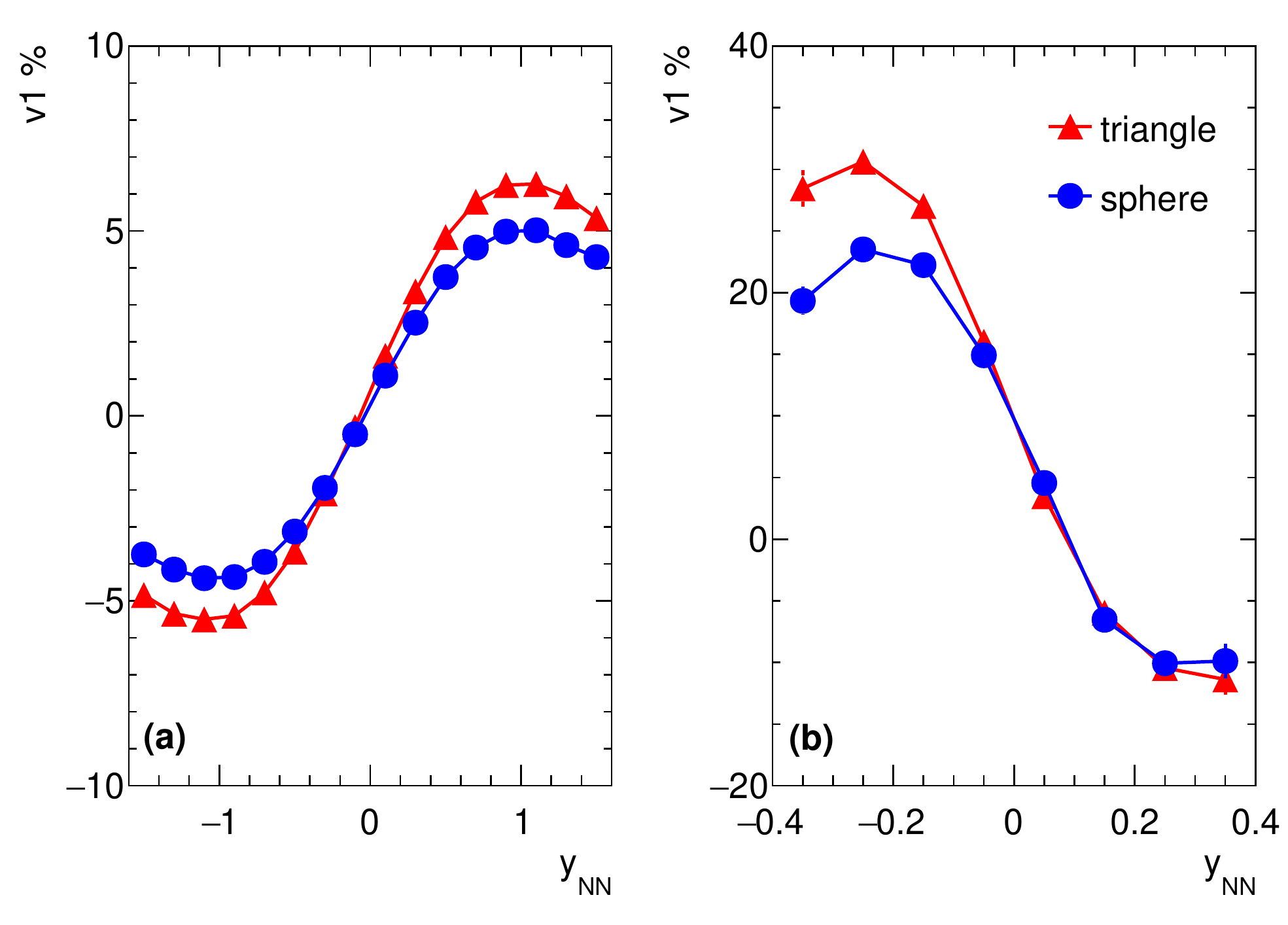}}
 \caption{The directed flows of photons (a) and particles (b) which were taken part in bremsstrahlung process. The marker is the same as figure \ref{flow}. }
\label{photon_proton}
\end{figure}

The elliptic flow of direct photons is shown in figure \ref{flow} (b). Photons tend to emit out of the reaction plane. And the difference of $v_2$ between two $^{12}$C configurations is observed. Comparing with triangle configuration case, the elliptic flow tends to emit out of plane in the sphere configuration of $^{12}$C case. As opposed to direct photons, the free protons tend to emit in plane as shown in panel (d). This $v_2$ anti-correlation between direct photons and free protons is in agreement with our previous results  \cite{LiuGH2008,LIU2008312,WangSS}.

In order to judge the relationship between direct photons and free nucleons more carefully. We compare the directed flow of $\gamma$ and particles taking part in the bremsstrahlung process in the figure \ref{photon_proton}. The original irregular curve is restored to $S$-shape, and a negative flow parameter is observed. It strongly indicates an attractive dominant interaction between nucleons in the early stage. Moreover, $v_1$ of triangular configuration is more intense, which is consistent with our analysis above. However, the curve is not fully symmetric, that is, it does not cross the origin, which indicates that the collective flow is influenced by the asymmetry of system in the early stage. After re-extraction, even in the case of asymmetry system, directed flow still keeps an anti-correlation relationship between direct photons and free protons. It should be emphasized again, the collective motion of direct photons is only a reflection of hadrons. However, its information will be washed out by surround nuclear matter intensely through the whole reaction. Fortunately, the hard photons as an alternative choice could take a unique undistorted information for reaction mechanism and geometric information of nuclei at the early stage to us. Another thing to note is that, we do not set an energy range of selected photons in the anisotropic flow calculation. It may cause asymmetry if we choose photons with an energy cut off.

\begin{figure}[htbp]
\resizebox{8.6cm}{!}{\includegraphics{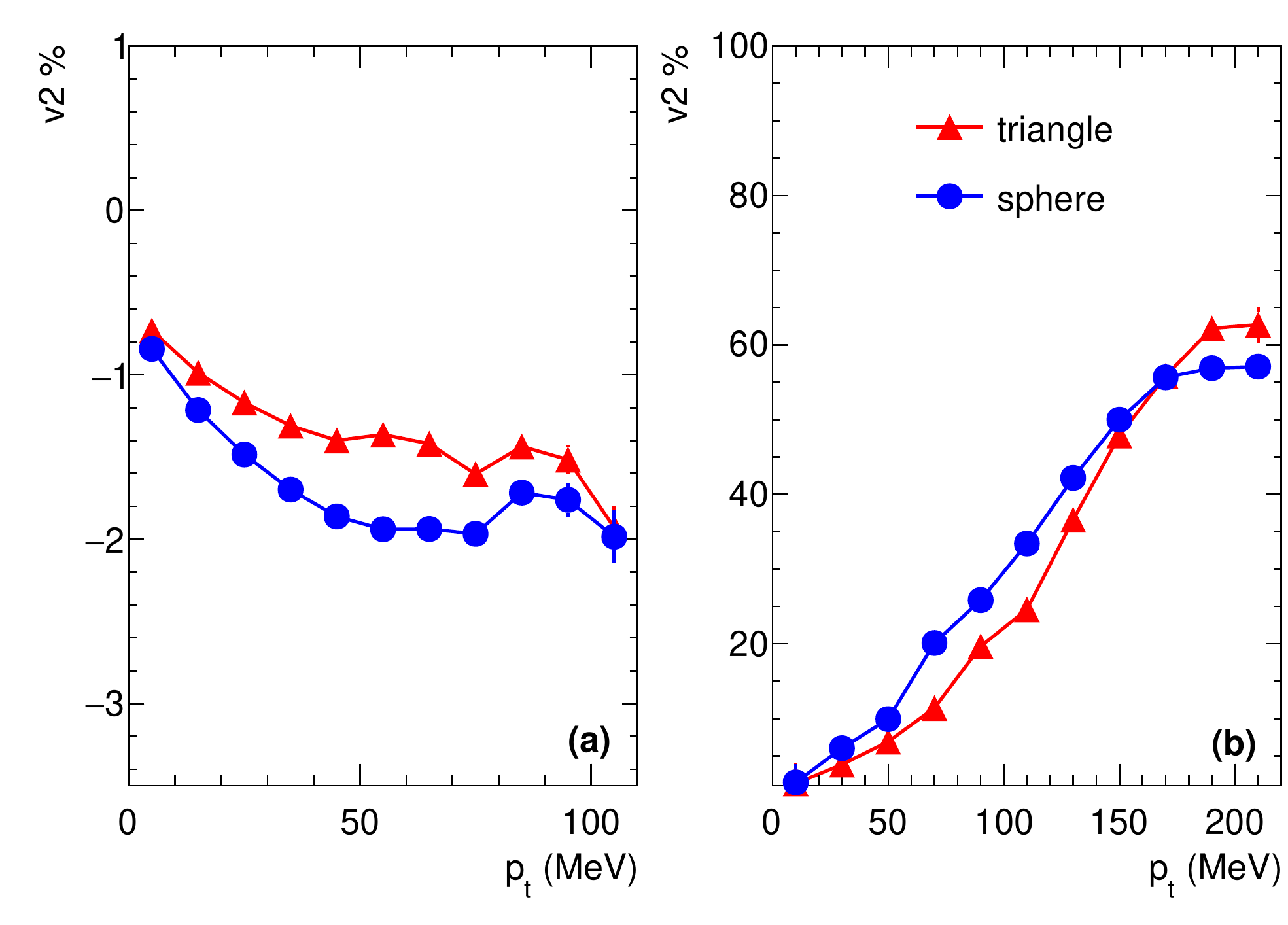}}
 \caption{The rapidity dependence of elliptic flows for direct photons (a) and free protons (b) selected with rapidity region $0.8y_{t}<y_{nn}<0.8y_{p}$. }
\label{flow_pt}
\end{figure}

The elliptic flows of direct photons and free protons as a function of rapidity is shown in figure \ref{flow_pt} (a) and (b), respectively. The free protons are selected with the cut of  $0.8y_{t}<y_{nn}<0.8y_{p}$. For free protons, it is found that a strong transverse momentum dependence of elliptic flow , and its value increases very fast with the transverse momentum in $p_t<$150 MeV. However, for direct photons, the transverse momentum dependence is much weaker. As the transverse momentum increases, it reaches saturation  quickly. Generally speaking, the free protons with high transverse momentum escape at earlier time and those with low transverse momentum experience more collisions. In other word, more frequent energy exchange occurs. Different from the situation of free protons, all direct photons come from the early stage. So their transverse momenta are not decided by the emitting time. It is a reason why the elliptic flow of direct photons shows a small transverse momentum dependence. Based on this analysis result, we deduce that an obvious discrepancy of collective flows is already formed at pre-equilibrium stage.

\section{Conclusion}
\label{sec:con}
In summary, we firstly obtain nice energy and angular spectra of direct photons  for  $^{86}$Kr + $^{12}$C at  $E/A$ = 44 MeV within our modified EQMD model. The magnitude and slope of hard photons is reasonably reproduced in comparison with experimental data. It indicates that our modified EQMD model can deal with bremsstrahlung process correctly within the dynamical wave packets.
Secondly, we investigate the collective flow of direct photons in peripheral collision. A nice $S$-shape curve of direct photons is obtained. In  contrast, the directed flow of free protons  loses much information due to the intense absorption by surround nuclear matter. Additional, difference of collective flows of direct photons were observed between triangle  and sphere configurations of $^{12}C$ which are initialized by the EQMD model self-consistently.
We also analyze the transverse momentum dependence of elliptic flows. It is found that the direct photons have a more moderate dependence on transverse momentum due to in the pre-equilibrium stage. Besides, it can be deduced that the discrepancy of collective flow caused by geometric effect is already appeared in the very early stage.
Totally, it is found that the direct photon as an untwisted observable can reflect the early information of system even in an asymmetry system. Besides, the present work sheds light on  an alternative probe to find a possible signal of $\alpha$-clustered structure in light nucleus by using direct photons.

This work was partially supported by the National Natural Science Foundation of China under Contract Nos. 11890710, 11890714 and 11961141003, the Strategic Priority Research Program of the CAS under Grants No. XDB34000000,  and Guangdong Major Project of Basic and Applied Basic Research No. 2020B0301030008.

\section{Appendix}
\label{appendix}

Strictly speaking, one should sample adequate times according to the Wigner function of a nucleon as follow:
\begin{equation}
\begin{aligned}
f_i({\mathbf{r}})=&\rho_{i}(\mathbf{r}) \\
=&\left(\frac{1}{\pi\lambda_i}\right)^\frac{3}{2}exp\left\{-\frac{1}{\lambda_i}\left(\mathbf{r-\mathbf{R}_i}\right)^2 \right\}\\
g_i({\mathbf{p}})=&\frac{w(\mathbf{r},\mathbf{p})}{\rho_{i}(\mathbf{r})}\\
=&\left(\frac{\lambda_i}{\pi\hbar^2}\right)^\frac{3}{2}exp\left\{-\frac{\lambda_i}{\hbar^2}\left[\mathbf{p}-\left(\mathbf{P}_i+\delta_i\hbar\mathbf{R}_i-\delta_i\hbar\mathbf{r}\right)\right]^2\right\}.
\end{aligned}\label{eq:sample_strictly}
\end{equation}

Here $f_i(\mathbf{r})$ means the probability of the $i$-th nucleon is at point $\mathbf{r}$, and the $g_i(\mathbf{p})$ represents the probability of this nucleon with momentum $\mathbf{p}$ when its position is known at $\mathbf{r}$. It is easily to get some quantities from Eq.\ref{eq:sample_strictly} as follow:

\begin{equation}
\begin{aligned}
\overline{\mathbf{r}}=&\mathbf{R}_i\\
\overline{\mathbf{r}^2}=&\mathbf{R}_i^2+\frac{3}{2}\lambda_i\\
\overline{\mathbf{p}}=&\mathbf{P}_i\\
\overline{\mathbf{p}^2}=&\overline{\left[\mathbf{P}_i-\delta_i\hbar\left(\mathbf{r}-\mathbf{R}_i\right)\right]^2}+\frac{3\hbar^2}{2\lambda_i}\\
=&\mathbf{P}_i^2+\overline{\delta_i\hbar\left(\mathbf{r}-\mathbf{R}_i\right)^2}+\frac{3\hbar^2}{2\lambda_i}\\
=&\mathbf{P}_i^2+\frac{3\hbar^2}{2\lambda_i}\left(1+\lambda_i^2\delta_i^2\right).
\end{aligned}
\end{equation}

If we take account the zero-pointer kinetic energy into momentum sampling, then our formulation looks like:
\begin{equation}
\begin{aligned}
\Delta\mathbf{p}=&\mathbf{p}-\mathbf{P}_i \\
\mathbf{p}_i=&\mathbf{P}_i+\sqrt{1-\frac{1}{M_i}}\times \Delta\mathbf{p}.
\end{aligned}
\end{equation}
Here $\mathbf{p}_i$ is the nucleon's momentum after its zero-point kinetic energy is taken into account.

If one samples coordinate and momentum adequately, it can be easily to prove:
\begin{widetext}
\begin{equation}
\begin{aligned}
\overline{\mathbf{p}_i^2}=&\overline{\left(\mathbf{P}_i+\sqrt{1-\frac{1}{M_i}}\times\Delta\mathbf{p}\right)^2}\\
=&\mathbf{P}_i^2+\left(1-\frac{1}{M_i}\right)\times\frac{3\hbar^2}{2\lambda_i}\left(1+\lambda_i^2\delta_i^2\right)\\
\overline{T_i}=&\frac{\mathbf{P}_i^2}{2m_i}+\left(1-\frac{1}{M_i}\right)\times\frac{3\hbar^2}{4m_i\lambda_i}\left(1+\lambda_i^2\delta_i^2\right)\\
=&\frac{\mathbf{P}_i^2}{2m_i}+\frac{3\hbar^2\left(1+\lambda_i^2\delta_i^2\right)}{4m_i\lambda_i}-T_\mathrm{zero}.
\end{aligned}
\end{equation}
\end{widetext}

So far, it proves that the kinetic energy obtained by this sampling method is consistent with the expected kinetic energy described by the Hamilton of the EQMD model. This part of kinetic energy is dominant in the namely "Fermi-motion" question in  paper of Maruyama et al \cite{EQMD}. In the actual calculation, what needs to be noted, there is appropriate simplification for simulation.

\bibliographystyle{plain}
\bibliography{mylib_flow}

\end{CJK*}
\end{document}